\documentstyle[11pt,aaspp4]{article}
\def\lsim{\la}
\def\gsim{\ga}
\usepackage{graphics}

\begin{document}

\title{\bf Constraints on Off-Axis X-Ray Emission from Beamed GRBs}
\author{Eric Woods$^1$ and Abraham Loeb$^2$}
\medskip
\affil{Astronomy Department, Harvard University, 60 Garden St.,
ambridge, MA 02138}
\altaffiltext{1}{email: ewoods@cfa.harvard.edu}
\altaffiltext{2}{email: aloeb@cfa.harvard.edu}

\begin{abstract}

We calculate the prompt x--ray emission as a function of viewing angle for
beamed Gamma--Ray Burst (GRB) sources. Prompt x-rays are inevitable due to
the less highly blueshifted photons emitted at angles $\gsim 1/\gamma$
relative to the beam symmetry axis, where $\gamma$ is the expansion Lorentz
factor.  The observed flux depends on the combinations $\gamma\Delta\theta$
and $\gamma\theta_v$, where $\Delta\theta$ is the beaming angle and
$\theta_v$ is the viewing angle.  We use the observed source counts of
$\gamma$--ray--selected GRBs to predict the minimum detection rate of
prompt x--ray bursts as a function of limiting sensitivity.  We compare our
predictions with the results from the {\it Ariel V} catalog of fast x--ray
transients, and find that {\it Ariel}'s sensitivity is not great enough to
place significant constraints on $\gamma$ and $\Delta\theta$.  We estimate
that a detector with fluence limit $\sim 10^{-7}~{\rm erg~cm^{-2}}$ in the
2--10 keV channel will be necessary to distinguish between geometries.
Because the x-ray emission is simultaneous with the GRB emission, our
predicted constraints do not involve any model assumptions about the
emission physics but simply follow from special-relativistic
considerations.

\end{abstract}

\keywords{gamma rays: bursts}

\section{Introduction}

With the 1996 launch of the {\it BeppoSAX} satellite (Boella et al. 1997),
it became possible to localize Gamma--Ray Burst (GRB) sources to arcminute
accuracy within a few hours after their detection.  Such localizations were
followed quickly by the discovery of delayed counterparts (``afterglows'')
in the x--ray (Costa et al. 1997), optical (van Paradijs et al. 1997), and
radio (Frail et al. 1997) regimes of the spectrum.  Subsequent
identifications of spectral features in the afterglows and host galaxies of
four GRBs led to the determination of source redshifts $z=0.835$ for GRB
970508 (Bloom et al. 1998), $z=0.966$ for GRB 980703 (Djorgovski et
al. 1998), $z=3.418$ for GRB 971214 (Kulkarni et al. 1998), and
$z=1.6$ for GRB 990123 (Bloom et al. 1999).
Thus the distance scale was firmly established as cosmological, and the
energy scale deduced from the observed fluences was very large, $\gsim
10^{51}$--$10^{54}$ ergs for isotropically--emitting sources.  The standard
picture which naturally follows from such an enormous energy release is a
relativistically--expanding fireball with a Lorentz-factor $\gamma\ga 10^2$
, which produces the observed gamma--rays with the formation of either
internal (Paczy\'nski \& Xu 1994; Rees \& M\'esz\'aros 1994; Sari \& Piran
1997; Pilla \& Loeb 1998) or external (e.g. M\'esz\'aros \& Rees 1993)
shocks.  Examples where the inferred energy release for isotropic emission
exceeds the expected radiative energy supply from compact stellar-mass
objects (Kulkarni et al. 1998; Fruchter 1998; Kulkarni et al. 1999),
suggest the existence of
beaming.

The degree of beaming in GRB sources is still poorly constrained.  The
expansion could be taking place in a jet whose opening half--angle is
anywhere from $\ll 1/\gamma$ (strongly beamed) to $\pi$ (isotropic
expansion).  Many of the current models for GRB progenitors involve systems
which have a rotation axis (e.g., Paczy\'nski 1998; Fryer et al. 1998;
Popham et al. 1998; Woosley 1998; M\'esz\'aros et al. 1998), possibly
leading to a preferred expansion direction for the explosion. With beaming,
the total energy necessary to produce the observed fluxes will be reduced
by a factor $\Delta\Omega/4\pi$, where $\Delta\Omega\ga 1/\gamma^2$ is the
solid angle subtended by the GRB radiation.  In addition, the total event
rate must be greater by a factor $4\pi/\Delta\Omega$.  As the Lorentz
factor of the emitting material declines with time, it is likely that the
delayed emission at longer wavelengths will be less relativistically beamed
than the jet collimation angle, and that detection rates of afterglows
(after correcting for different detector sensitivities) will be
correspondingly larger (Rhoads 1997; Perna \& Loeb 1998).  Grindlay (1999)
has investigated this possibility by comparing the discovery rate of GRBs
by the {\it BATSE} experiment with that of fast x--ray transient sources
(those which exceeded the flux threshold for only $\sim 1$ orbit) by the
{\it Ariel V} satellite (Pye \& McHardy 1983) two decades ago; he found
that the rates were roughly consistent with no differential beaming between
the $\gamma$--ray and the x--ray emission. The naive interpretation of this
constraint is, however, sensitive to the assumption that each GRB is
followed by a bright X-ray afterglow. In this paper, we try to avoid this
assumption and consider instead the inevitable (simultaneous) emission of
X-rays at an angle relative to the GRB beam, purely due to special
relativistic considerations.

If the $\gamma$--ray emission is beamed, it is unavoidable that there will
be some photons which are detectable even when the observer is not within
the $\gamma$--ray emitting beam.  These are the photons emitted at large
angles to the expansion direction by material near the edge of the jet.
Such photons will not generally be detectable as $\gamma$--rays; since they
are less highly blueshifted than forward--emitted photons, they will be
seen in the x--ray band.  Thus, there should be some minimum degree of
differential beaming between the $\gamma$--rays and the {\it prompt}
x--rays.  The degree to which this will affect the x--ray detection rate
depends on the expansion Lorentz factor $\gamma$ and the beaming
half--angle $\Delta\theta$; comparison of the predicted rate with that
observed by {\it Ariel V} can yield constraints on $\gamma$ and
$\Delta\theta$.

In this paper, we use the {\it BATSE} source counts and a simple
geometrical model for the GRB sources to predict the prompt x--ray
detection rate as a function of limiting flux.  In \S 2, we derive the
expression for the flux as a function of viewing angle for a given
geometry.  In \S 3, we calculate the x--ray detection rate for various
values of $\gamma$ and $\Delta\theta$, and compare the results with the
rate observed by {\it Ariel V}.  In \S 4, we present our conclusions.

\section{Calculation of the X--Ray Flux}

\subsection{Emission from Optically--Thin, Relativistically--Moving Material: General Case}

It will be instructive first to calculate the observed flux for the general
case of emission from optically thin, relativistically--moving material.
In doing so, we closely follow the analysis of Granot, Piran, \& Sari
(1998).  Suppose we have some optically--thin material with rest--frame
emissivity $j'_{\nu'}$, measured in
${\rm erg~s^{-1}~cm^{-3}~Hz^{-1}~sr^{-1}}$. 
Following from considerations
of Lorentz--invariance (Rybicki \& Lightman 1979), the lab--frame
emissivity per frequency $\nu$ is given by
\begin{equation}
j_\nu = \frac{j'_{\nu'}}{\gamma^2(1-\beta\mu_v)^2},
\label{jnu}
\end{equation}
where $\gamma=(1-\beta^2)^{-1/2}$ is the local Lorentz factor, $\mu_v$ is
the cosine of the angle between the local velocity and the direction to the
detector, and $\nu'=\nu\gamma(1-\beta\mu_v)$.  In general, $j'_{\nu'}$ is a
function of position, time, frequency, and direction; we are only concerned
with photons which are emitted in the direction $\Omega'_d$ toward the
detector.

Let us use a spherical coordinate system $\vec{r}=(r,\theta,\phi)$, where
the coordinates are measured in the lab frame; let the $\theta =0$ axis
($z$--axis) point toward the detector.  Suppose the detector is at a
distance $D$ from the source at the origin (see Figure 1).  Furthermore,
let $\alpha$ be the angle that a given ray makes with the normal to the
detector.  The flux at the detector, measured in
${\rm erg~s^{-1}~cm^{-2}~Hz^{-1}}$, is then given by
\begin{equation}
F_\nu = \int I_\nu(\alpha,\phi)~d\Omega = \int_0^{2\pi}d\phi \int_0^{\pi} d\alpha~\sin\alpha~
\cos\alpha~I_\nu(\alpha,\phi),
\label{fluxint}
\end{equation}
where $I_\nu(\alpha,\phi)$ is the intensity along a ray incident on the detector in direction
$(\alpha,\phi)$.  We will only consider situations where the source is very far away, so that all
rays come in very nearly normal to the detector.  Note that in this case, $\alpha\ll 1$, so 
$\sin\alpha\approx\alpha$ and $\cos\alpha\approx 1$; furthermore, the only significant
contribution to the integral comes
from $\alpha\lsim L/D\ll 1$, where $L$ is the size of the source.  In the approximation that
$\alpha\ll 1$, simple geometry gives
\begin{equation}
\alpha = \frac{r}{D}\sin\theta = \frac{r}{D} \sqrt{1-\mu^2},
\label{alpha}
\end{equation}
where we have defined $\mu\equiv\cos\theta$.  The equation of a ray incident at angle $\alpha$ is thus
\begin{equation}
r = \frac{\alpha D}{\sin\theta},
\end{equation}
so the change in $r$ corresponding to an increment in $\theta$ along the ray is given by
\begin{equation}
dr = -\alpha D \frac{\cos\theta}{\sin^2\theta} d\theta.
\end{equation}
Therefore, the line element along a ray is given by
\begin{equation}
ds = (r^2 d\theta^2 + dr^2)^{1/2} = \frac{\alpha D}{(1-\mu^2)^{3/2}} d\mu.
\label{ds}
\end{equation}
For an optically--thin medium, the contribution to the intensity from a ray
segment $ds$ is $dI_\nu = j_\nu ds$.  Note that if this $dI_\nu$ is
received at the detector at time $T$, then it relates to the emissivity
$j_\nu$ at an {\it earlier} time $t$, due to the light--travel time.  Let
$T=0$ be the arrival time at the detector of a photon emitted at the origin
at $t=0$; inspection of the geometry in Figure 1 gives
\begin{equation}
t = T + \frac{r\mu}{c}.
\label{time}
\end{equation}
Combining equations (\ref{jnu}), (\ref{fluxint}), (\ref{ds}), and (\ref{time}) gives
\begin{equation}
F_\nu(T) = D \int_0^{2\pi} d\phi \int_0^{\alpha_m} \alpha^2 d\alpha
\int_{-1}^1 d\mu \frac{j'_{\nu\gamma(1-\beta\mu_v)}(\Omega'_d,\vec{r},T+r\mu/c)}
{\gamma^2 (1-\beta\mu_v)^2 (1-\mu^2)^{3/2}}.
\label{fluxgen}
\end{equation}
Here, $\alpha_m=L/D$ where $L$ is the projected size of the emitting region;
$\Omega'_d$ is the direction toward the detector, as measured in the rest frame;
and $\vec{r}=(r,\theta,\phi)=(\alpha D/\sqrt{1-\mu^2},\cos^{-1}\mu,\phi)$
is the lab--frame position vector.

Equation (\ref{fluxgen}) is quite general; if we can specify the rest--frame emissivity
$j'_{\nu'}(\Omega'_d,\vec{r},t)$, perhaps from some physical model of the radiative
processes taking place, then we can calculate the flux at the detector.  Note that
in general, $\gamma$ (and therefore $\beta$) will be functions of $\vec{r}$ and $t$;
this dependence can often be derived from hydrodynamical considerations.

In the case of radial expansion, $\mu_v=\mu$, and we may switch integration
variables from $\mu$ to $\nu'$, and rewrite equation
(\ref{fluxgen}) as
\begin{equation}
F_\nu(T) = \frac{\nu D}{\gamma\beta} \int_0^{2\pi} d\phi \int_0^{\alpha_m}
\alpha^2 d\alpha \int_{\nu\gamma(1-\beta)}^{\nu\gamma(1+\beta)}
\frac{d\nu'}{{\nu'}^2}
~\frac{j'_{\nu'}[\Omega'_d,\vec{r},T+(1-\frac{\nu'}{\gamma\nu})\frac{r}{\beta c}]}
{\{1-[\frac{1}{\beta}(1-\frac{\nu'}{\gamma\nu})]^2\}^{3/2}}.
\label{fluxrad}
\end{equation}

\subsection{GRB Emission from a Conical Section of an Expanding Spherical 
Shell}

Let us assume that the expanding material is confined to a cone of opening
half--angle $\Delta\theta$, and that the emission takes place in a thin
shell which moves radially outward at the expansion speed $\beta c$, with
associated Lorentz factor $\gamma$.  Realistically, based on hydrodynamical
considerations, the shell should be of radial thickness $\sim 0.1\beta ct$,
but our results are not changed significantly by assuming that it is
infinitesimally thin.

The observed spectral shape of GRBs is well--reproduced by a broken power
law.  Usually, the two segments with different power--law indices are
joined smoothly with an exponential transition (Band et al. 1993).
Recently, Ryde (1998) devised a single analytical function which also
produces a smoothly--broken power law. We adopt this functional form for
the frequency dependence of the rest--frame emissivity, since it yields a
spectral shape similar to that observed (with the same form but with a
blue-shifted break frequency and a broader power--law transition).  We
assume typical power--law indices, $F_\nu'\propto(\nu')^0$ for low
frequencies and $F_\nu'\propto(\nu')^{-1}$ for high frequencies (Band et
al. 1993). In the case where the axis of the jet points directly at the
observer, we have
\begin{equation}
j'_{\nu'}(\vec{r},t) = A(t)~\delta(r-\beta ct)~H(\Delta\theta-\theta)~
\left[1+\left(\frac{\nu'}{\nu'_0}\right)^{\frac{2}{\eta}}\right]
^{-\frac{\eta}{2}}.
\label{jsphere}
\end{equation}
Here, $A(t)$ is the time--dependent normalization, and $H(x)$ is the
Heaviside step function, $\nu'_0$ is the transition frequency, and
$\eta=\ln(1+\Delta\nu'/\nu'_0)$, where $\Delta\nu'$ is the width of the
power--law transition.  In order to achieve agreement with the results of
Strohmayer et al. (1998) for the average ratio of x--ray to $\gamma$--ray
flux of {\it Ginga}--detected bursts, $F(2$--$10~{\rm
keV})/F(50$--$300~{\rm keV})=0.24$, we require that the observed spectral
break occur at $h\nu_0=18$ keV.  This is low compared to the typical break
energy $\sim 150$ keV for BATSE bursts (Band et al. 1993), since {\it
Ginga} was sensitive to lower--energy photons.  Note that even the BATSE
bursts span a range of break energies (10--$10^3$ keV); and since {\it
Ginga} is one of the few satellites to have probed the prompt x--ray
emission from GRBs, we adopt the {\it Ginga} value.  Given that for a
typical GRB expanding with Lorentz factor $\gamma\gg 1$ toward the
observer, the rest--frame frequency corresponding to $\nu_0$ is
$\nu'_0\approx\nu_0/(2\gamma)$, we use $h\nu'_0=9\gamma^{-1}$ keV.  The
width $\Delta\nu'$ is unimportant for our analysis, since it does not
appreciably affect the fluxes we calculate; we assume $h\Delta\nu'=1$ keV.
The observed GRB spectral shape predicted from equation (\ref{jsphere})
will have the same form, but with the break frequency $\nu_0$ blueshifted,
and the transition width $\Delta \nu$ broadened.  Of course, in general,
$\nu'_0$ will be a function of time; for simplicity, we assume that it is a
constant (i.e. constant spectral shape at the source).  Note also that
$\nu'j_{\nu'}\propto{\rm const}$ for $\nu'\gg\nu'_0$, which means that we
need to introduce a high--end cutoff frequency $\nu'_c$ so that $j'_{\nu'}$
will not possess a logarithmic ultraviolet divergence.  We set this upper
cutoff to correspond to an observed photon energy of 5 MeV.

Recalling that $\nu'=\nu\gamma(1-\beta\cos\theta)$, we may combine equations
(\ref{fluxrad}) and (\ref{jsphere}) to obtain
\begin{equation}
F_\nu(T) = \frac{2\pi\nu D}{\gamma\beta} \int_0^{\alpha_m} \alpha^2~d\alpha
\int_{\nu\gamma(1-\beta)}^{\nu\gamma(1-\beta\cos\Delta\theta)} \frac{d\nu'}{{\nu'}^2} \left[1+\left(\frac{\nu'}{\nu'_0}\right)
^{\frac{2}{\eta}}\right]^{-\frac{\eta}{2}}
\frac{\delta\left(\frac{\nu'r}{\gamma\nu}-\beta cT
\right) A\left(\frac{\gamma\nu}{\nu'}T\right)}
{\{1-[\frac{1}{\beta}(1-\frac{\nu'}{\gamma\nu})]^2\}^{3/2}},
\label{fluxnu}
\end{equation}
where $\alpha_m=L/D$ is the angular size of the source when $A(t)$
effectively cuts off, and equation (\ref{alpha}) gives the relation between
$r$ and $\alpha$ with $\mu=\beta^{-1}[1-(\nu'/\nu\gamma)]$.  
Note that equation (\ref{fluxnu}) implies that the observed effective
duration of the event is inversely proportional to the observed frequency.
This is due to the longer travel time for the less--blueshifted photons
coming in from off the line of sight.

Now, recall the following property of the Dirac delta function: for $f(y)$
monotonic,
\begin{equation}
\delta[f(y)] = \left|\frac{df}{dy}\right|_{y_0}^{-1} \delta(y-y_0),
\end{equation}
where $y_0$ is the zero of $f(y)$.  This allows us to transform the
delta function in equation (\ref{fluxnu}) to a delta function in $\alpha$. 
Thus, we can perform 
the $\alpha$--integration in equation (\ref{fluxnu}) to obtain
\begin{equation}
F_\nu(T) = 2\pi\beta\gamma^2 \left(\frac{cT}{D}\right)^2 \nu^4
\int_{\nu\gamma(1-\beta)}^{\nu\gamma(1-\beta\cos\Delta\theta)}
\left[1+\left(\frac{\nu'}{\nu'_0}\right)^{\frac{2}{\eta}}\right]
^{-\frac{\eta}{2}} A\left(\frac{\gamma\nu}{\nu'}T\right)
\frac{d\nu'}{{\nu'}^5}.
\label{fluxon}
\end{equation}
The integral may
be computed numerically for any observed frequency $\nu$ and time $T$.  The
factor $(cT/D)^2$ reflects the increase in angular size of the shell as it
expands.  The case $\Delta\theta=\pi$ corresponds to spherical expansion.
For simplicity, we adopt the form $A(t)=A_0\exp(-t/\tau)$; this produces
a single--pulse lightcurve, with total fluence
\begin{equation}
S_\nu = \int_0^\infty F_\nu(T) dT =
\frac{4\pi\beta A_0\tau\nu}{\gamma} \left(\frac{c\tau}{D}\right)^2 
\int_{\nu\gamma(1-\beta)}^{\nu\gamma(1-\beta\cos\Delta\theta)}
\left[1+\left(\frac{\nu'}{\nu'_0}\right)^{\frac{2}{\eta}}\right]
^{-\frac{\eta}{2}} \frac{d\nu'}{{\nu'}^2}.
\label{fluenceon}
\end{equation}

In the case where the line of sight to the observer $(\theta=0)$ lies
entirely outside the cone, the geometry gets a bit more complicated.  Let
$\theta_v$ be the angle between the direction to the detector and the axis
of the emission cone (see Figure 1).  Without loss of generality, we may
also place the axis of the cone at $\phi=0$.  In this case, the law of
cosines for spherical triangles (e.g., Peebles 1993) yields the following
equation in $(\theta,\phi)$ coordinates for the boundary of the cone:
\begin{equation}
\cos\theta_v\cos\theta + \sin\theta_v\sin\theta\cos\phi = \cos\Delta\theta.
\label{cosines}
\end{equation}
This translates into a more complicated angular dependence in equation
(\ref{jsphere}).  We now have that $j'_{\nu'}$ is non--zero for
$|\theta-\theta_v|<\Delta\theta$; for a given $\theta$, the limits on
$\phi$ are determined by solving equation (\ref{cosines}).  Thus, on the
right--hand side of equation(\ref{jsphere}), we have instead of
$H(\Delta\theta-\theta)$:
\begin{equation}
H(\Delta\theta-|\theta-\theta_v|)
~H\left[\cos\phi - \left(\frac{\cos\Delta\theta-\cos\theta_v\cos\theta}
{\sin\theta_v\sin\theta}\right)\right].
\end{equation}
We then obtain for the fluence
\begin{equation}
S_\nu = \frac{4\beta A_0\tau\nu}{\gamma} \left(\frac{c\tau}{D}\right)^2 
\int_{\nu\gamma[1-\beta\cos(\theta_v-\Delta\theta)]}
^{\nu\gamma[1-\beta\cos(\theta_v+\Delta\theta)]}
\Delta\phi(\nu') 
\left[1+\left(\frac{\nu'}{\nu'_0}\right)^{\frac{2}{\eta}}\right]
^{-\frac{\eta}{2}} \frac{d\nu'}{{\nu'}^2},
\label{fluxoff}
\end{equation}
where solving from equation (\ref{cosines}) with
$\cos\theta=\mu=\beta^{-1}(1-\nu'/\gamma\nu)$ yields
\begin{equation}
\Delta\phi(\nu') = \cos^{-1}\left\{\frac{\cos\Delta\theta-\beta^{-1}(1-\nu'/\gamma\nu)\cos\theta_v}
{\sin\theta_v\sqrt{1-[\beta^{-1}(1-\nu'/\gamma\nu)]^2}}\right\}.
\end{equation}
This is the same result as equation (\ref{fluenceon}), except that the limits of $\nu'$--integration
have changed, and the $\phi$--integration now gives $2\Delta\phi$ instead of $2\pi$.

In the intermediate case, where the observer is inside the cone but
off-axis, the flux is given by the sum of the right--hand sides of
equations (\ref{fluxon}) and (\ref{fluxoff}), with the upper limit of
integration in (\ref{fluenceon}) replaced by
$\nu\gamma[1-\beta\cos(\theta_v-\Delta\theta)]$.

\section{Calculation of Prompt X-Ray Source Counts}

\subsection{Derivation the X--Ray Rate from the Gamma-Ray Rate}

We would like to calculate the expected detection rate of x--ray transients
as a function of limiting fluence, due to prompt emission from GRBs.  In
particular, we consider the energy channel of 2--10 keV probed by {\it
Ariel V} (Pye \& McHardy 1983), and compute the rate of X-ray transients
based on the GRB source counts found by {\it BATSE} in the 30--2000 keV
band (Bloom, Fenimore, \& in 't Zand 1996).  We use fluence instead of flux
because the {\it Ariel V} data have very coarse time resolution; the
threshold is based on the number of counts detected by the satellite over
one orbit $(\sim 100~{\rm min}= 6\times10^3~{\rm s})$, and is reported as
$S_x=2.4\times 10^{-6}~{\rm erg~cm}^{-2}$ in the 2--10 keV range (Pye \&
McHardy 1983; Grindlay 1999).  Our analysis yields the result that the
observed duration at x--ray frequencies goes roughly like $\tau_{\rm
x}\sim\tau_\gamma(\nu_\gamma/\nu_{\rm x})$, where $\tau_\gamma$ is the
duration as observed in gamma rays, and $\nu_{\rm x}$ and $\nu_\gamma$ are
the effective observed frequencies in x--rays and gamma rays respectively.
Since typically $\tau_\gamma\la 10$ s, we have that $\tau_{\rm x} \la 10^3$
s.  Thus, the integration time for {\it Ariel V} was sufficiently long for
it to detect the entire fluence of the prompt emission.

If the gamma--ray emission is beamed into a cone of half--angle
$\Delta\theta$, then we will only detect a GRB source in gamma--rays if our
line of sight falls within an angle $\sim 1/\gamma$ of the edge of the
beam.  Since the viewing angles $\theta_v$ should be distributed
isotropically; it follows that the fraction of sources whose viewing angles
are in the range $(\theta_1,\theta_2)$ is $(\cos\theta_1-\cos\theta_2)/2$.
Thus, the actual event rate is enhanced over the detected event rate by a
factor
\begin{equation}
\frac{4\pi}{\Delta\Omega} = \frac{2}{1-\cos[\max(\Delta\theta,1/\gamma)]}.
\end{equation}

Note that equations (\ref{fluxon}) and (\ref{fluxoff}) allow one to
calculate the ratio of the {\it Ariel V} x--ray fluence $S_x$ to the
{\it BATSE} gamma--ray fluence $S_\gamma$:
\begin{equation}
\frac{S_x}{S_\gamma} = \frac{\int_2^{10}S_\nu(\theta_v)d\nu}
{\int_{30}^{2000}S_\nu(\theta_v=0)d\nu},
\label{ratio}
\end{equation}
where the frequency $\nu$ is in units of keV.

The number of {\it Ariel V} x--ray sources with fluences brighter than $S_x$ is
\begin{equation}
N(>S_x) = \frac{1}{1-\cos[\max(\Delta\theta,1/\gamma)]} 
~\int_0^\infty\{1-\cos[\theta_v(S_x/S_\gamma)]\}
~n(S_\gamma)dS_\gamma,
\label{counts}
\end{equation}
where $n(S_\gamma)dS_\gamma$ is the number of sources with fluences in the
range $(S_\gamma,S_\gamma+dS_\gamma)$ seen by {\it BATSE}, and $\theta_v$
is the solution for $\theta_v$ in equation (\ref{ratio}), since 
$S_x/S_\gamma$ is a monotonic function of $\theta_v$ (cf. Fig. 2).

\subsection{Results}

We apply equations (\ref{ratio}) and (\ref{counts}), with the form of
$F_\nu(T)$ derived in \S 2, to predict the source detection rate in the
x--ray (2--10 keV) range probed by {\it Ariel V}.  We find that the results
depend only on the combinations $\gamma\theta_v$ and $\gamma\Delta\theta$,
as expected from ultra--relativistic beaming.  We calculate the source counts
for four beaming angles: $\gamma\Delta\theta=0.1$, 1, 10, and 100.

In Figure 2, we plot the x--ray fluence as a function of viewing angle
$\theta_v$.  Clearly, for a given event the overall normalization depends
on a measurement of, say, the {\it BATSE} gamma--ray fluence, so we plot
the ratio $S_x(\theta_v)/S_\gamma(0)$ [cf. Eq.~(\ref{ratio})].  For
comparison, we also show the gamma--ray fluence as a function of viewing
angle, plotted as the ratio $S_\gamma(\theta_v)/S_\gamma(0)$.  Note that
the gamma--ray fluence generally falls off to within an angle $\gamma^{-1}$
away from the beam's edge, as expected from special relativity.  The x--ray
fluence decreases somewhat more gradually with viewing angle; this effect
is significant out to a few times $\gamma^{-1}$ away from the beam's edge.
This is due to the relative redshifting of photons at larger viewing
angles, and is the basis for our assertion that for strongly beamed bursts
(to within $\sim\gamma^{-1}$), we should see more prompt transient events
in x--ray searches than in gamma--ray searches.  The next step is to apply
Equation (\ref{counts}) to see how sensitively the detection rate should
depend on the beaming angle.

In Figure 3, we plot the source detection rate as a function of limiting
flux [cf. Equation (\ref{counts})].  We see that
the counts only differ appreciably for limiting fluences fainter than a few
times $10^{-7}~{\rm erg~cm^{-2}}$.  For comparison, we display the {\it
Ariel V} data point, with $1\sigma$ Poisson errors. 
The {\it Ariel} data are clearly marginally consistent at the $1\sigma$
level with any degree of beaming, and we would need an experiment at
least an order of magnitude more sensitive to be able to place meaningful
constraints on the beaming.  Finally, we note that the results for
$\gamma\Delta\theta\sim 100$ agree with the $\log N-\log S$ plots obtained
from {\it BATSE} (Bloom, Fenimore, \& in 't Zand 1996), once one corrects
for the frequency band.

\section{Conclusions}

For a given GRB spectral shape, the predicted number of X-ray transients
with no GRB counterparts depends only on the combination
$\gamma\Delta\theta$, where $\Delta\theta$ is the beaming angle and
$\gamma$ is the expansion Lorentz factor of the sources.  The excess of
X--ray transients in the 2--10 keV band becomes significant only at
fluences $S_x\lsim 10^{-6}~{\rm erg~cm^{-1}}$; therefore, the sample of
fast x--ray transients detected by the {\it Ariel V} satellite in the
1970's (with sensitivity $2.4\times 10^{-6}~{\rm erg~cm^{-1}}$) is not deep
enough to place model-independent constraints on the beaming angle.  It
will require more sensitive instruments, such as the {\it MAXI} all--sky
monitor (Kawai et al. 1996)
aboard the soon--to--be--launched {\it International Space
Station}, or {\it MOXE} on {\it Spectrum-X} (in 't Zand, Priedhorsky,
\& Moss 1994), with a comparable
or larger sample size, to rule out strong beaming at the $2\sigma$--level.  In
addition, we may hope to use this analysis in the near future to learn more
about the low--frequency spectral index in GRBs.

In contrast with previous limits (e.g., Grindlay 1999; Perna \& Loeb 1998),
our analysis is not sensitive to any model assumptions about the physics of
the GRB/afterglow emission, but rely only on special relativistic
considerations and the observed properties of GRBs.  It is {\it inevitable}
that strongly--beamed GRBs will be accompanied by a larger number of
detectable prompt x--ray events; it is only a matter of pushing the
sensitivity limits down by an order of magnitude or so before we might
expect to observe this excess.

\acknowledgements

This work was supported in part by the NASA grants NAG5-7768 and NAG5-7039.

\vfill\eject







\vfill\eject
\begin{figure}
\plotone{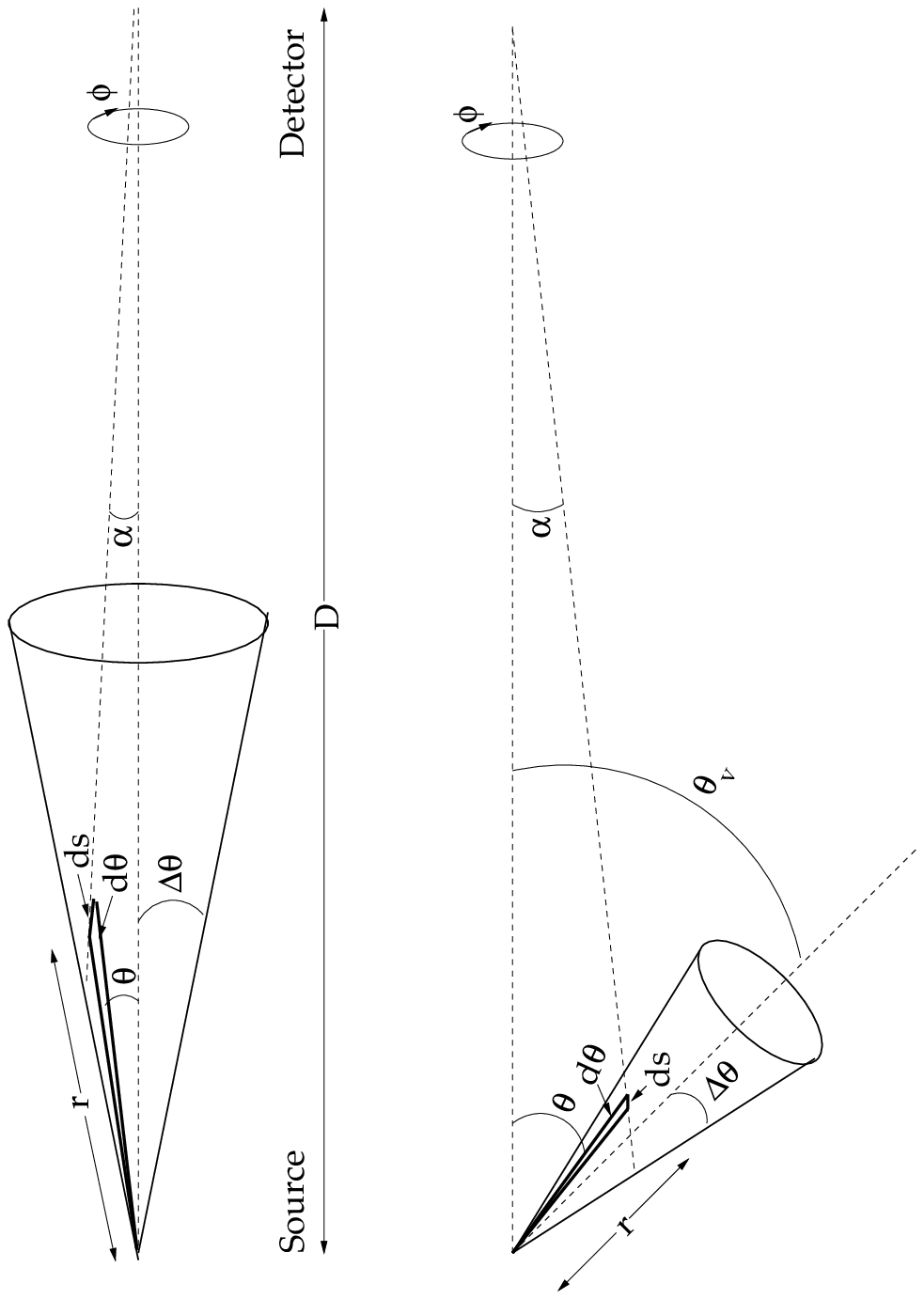}
\caption{Geometry for calculating the flux received from
a relativistically expanding source.  Top panel shows the case
of a jet aligned with the line of sight; bottom panel shows the
case of an off-axis jet.}
\end{figure}

\begin{figure}
\plotone{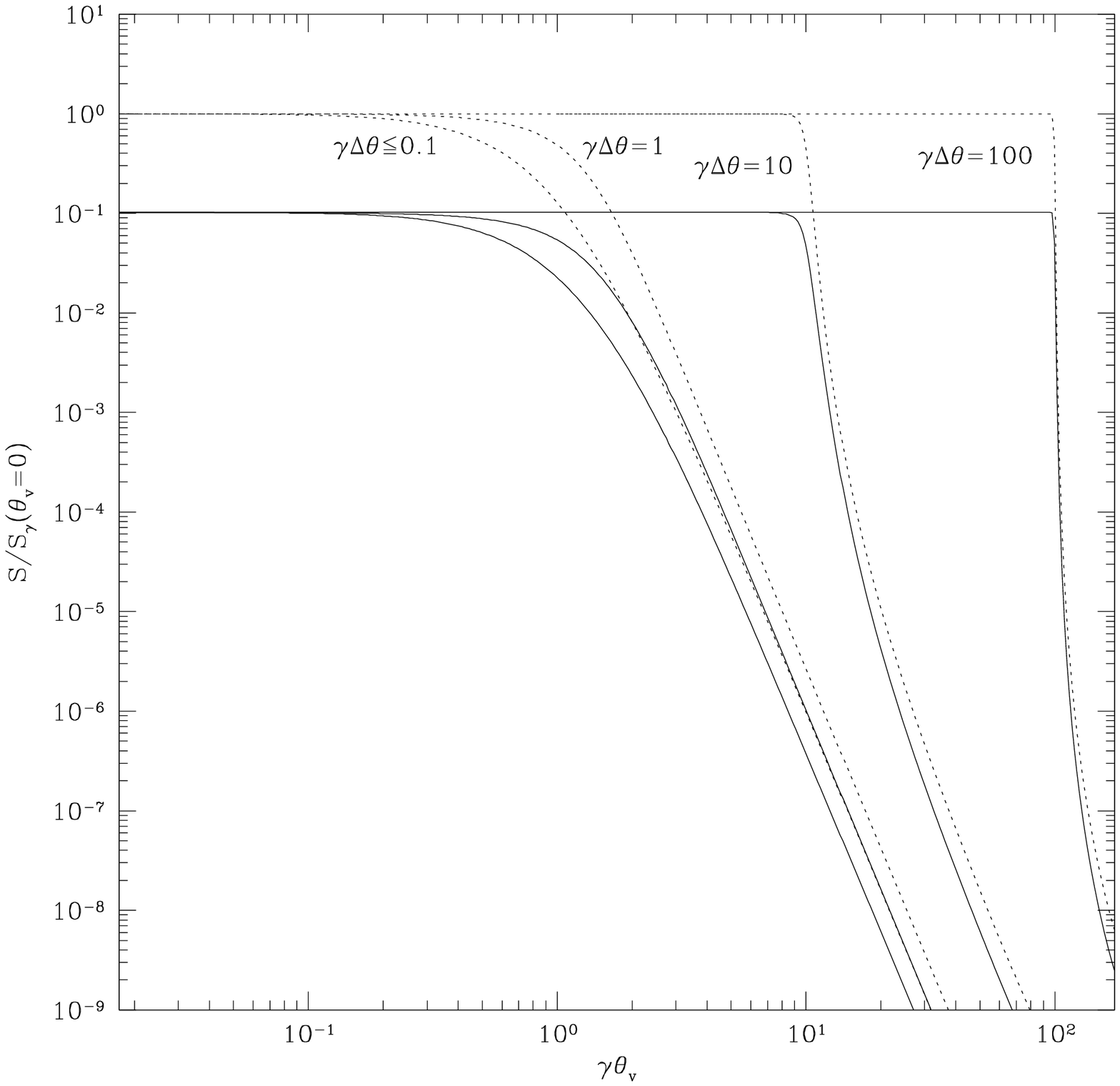}
\caption{X--ray fluence $S_x$ (solid curves) and gamma--ray
fluence $S_\gamma$ (dotted curves) as a function of viewing angle
$\theta_v$, in units of the on--axis $\gamma$--ray fluence,
$S_\gamma(0)$.  From
left to right, $\gamma\Delta\theta=0.1$, 1, 10, and 100.}
\end{figure}

\begin{figure}
\plotone{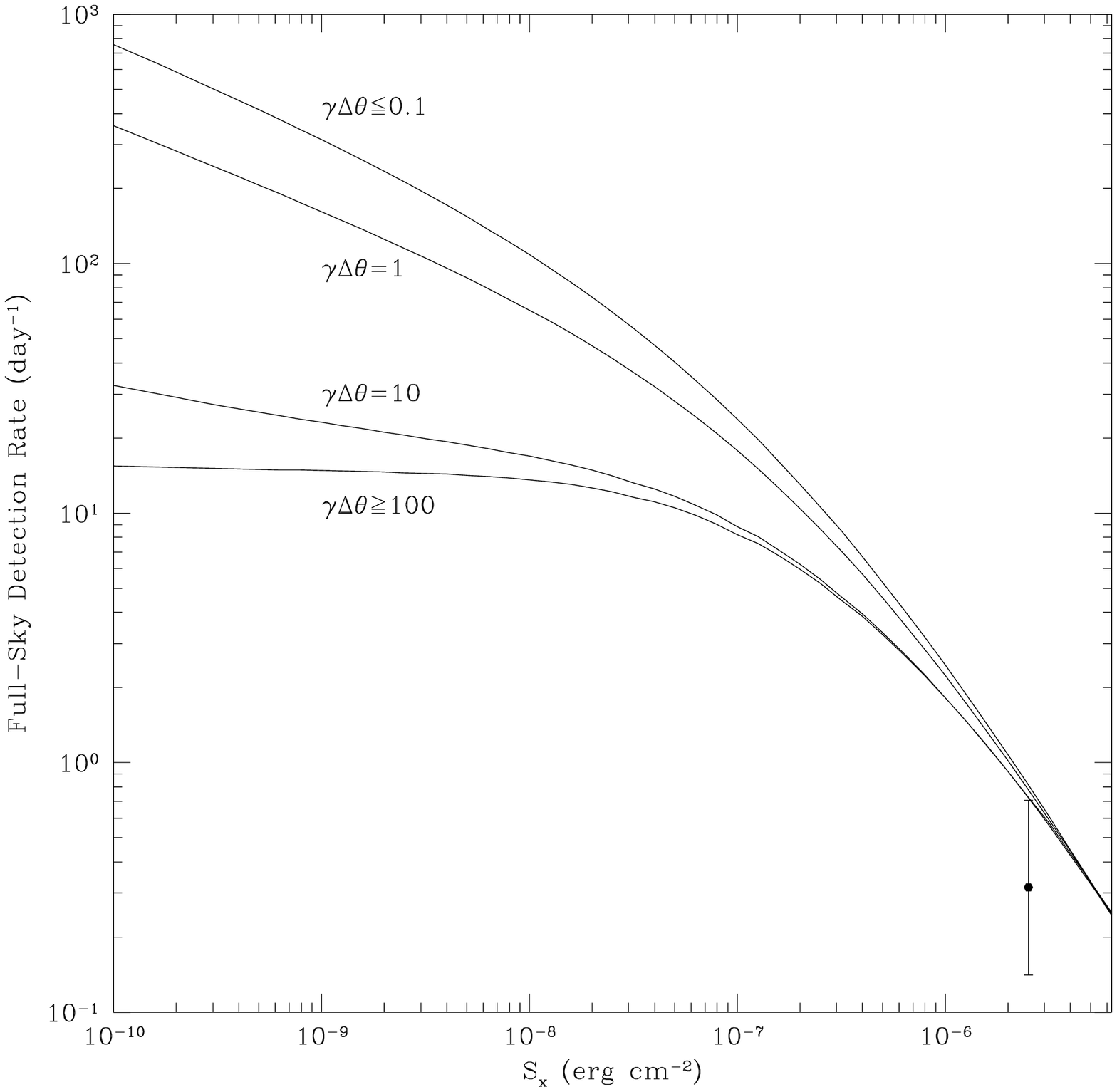}
\caption{X--ray (2--10 keV) source detection rate as a function of
limiting fluence, for $\gamma\Delta\theta=0.1$, 1, 10, and 100. 
The data point is the rate of fast x--ray transient
detection by {\it Ariel V}, which had a sensitivity of $S_x=2.4\times
10^{-6}~{\rm erg~cm^{-2}}$ in the 2--10 keV channel.  The error bar is
based on the estimated Poisson uncertainty.}
\end{figure}

\end{document}